\documentclass[twocolumn,showpacs,prb,amsmath,amssymb]{revtex4}

\usepackage{graphicx}
\usepackage{bm}

\begin{document}

\title{Kohn-Luttinger pseudo-pairing in a two-dimensional Fermi-liquid}

\author{V. M. Galitski}
\author{S. \surname{Das Sarma}}
\affiliation{Condensed Matter Theory Center and Center for
Superconductivity Research, 
Department of Physics, University of Maryland, College Park, Maryland
20742-4111}

\begin{abstract}
We consider possible superconducting instabilities in 
a two-dimensional Fermi system with short-ranged repulsive interactions
between electrons. The possibility of an unusual superconducting
paring due to the Kohn-Luttinger mechanism is examined. 
The quasiparticle scattering amplitude
 is shown to possess an attractive harmonic in 
second-order perturbation theory for finite values of
the energy transfer. The corresponding singularity in the pairing vertex
leads to a superconducting pairing of the electron excitations with finite energies.
We identify the energy transfer in the Cooper channel 
 as the binding energy of the excited pair. At low enough temperatures, the Fermi system is a mixture 
 of normal electron excitations and fluctuating $d$-wave Cooper pairs
 possessing a finite gap. 
\end{abstract}


\maketitle


\section{Introduction}

Superconductivity induced by mechanisms other than electron-phonon interactions
has been of long-standing interest. 
Throughout the last decade there has been continuing theoretical search for unconventional
superconductivity mechanisms, particularly in two-dimensional systems. 
This interest has been, indeed, motivated by novel
superconducting materials such as high-$T_c$ cuprates, organic superconductors 
as well as by the studies of $\vphantom{H}^3{\rm He}$ films. Currently, there is no full
understanding of the physical processes responsible for the pairing in those systems.

Kohn-Luttinger effect is one of the oldest as well as among the most appealing and elegant physical 
effects, which
might be considered within this quest. Back in 1965, Kohn and Luttinger \cite{KL} showed that 
any three-dimensional electron system with repulsive interactions between particles
was unstable against a superconducting transition at extremely low temperatures.
The origin of the effect is that the screening of the bare interaction leads to 
the well-known Friedel oscillations in the electron density and to similar oscillations
in the scattering amplitude. The renormalized interaction acquires a long-ranged oscillatory component.
Thus, there appear some regions where the effective interaction is attractive. This leads
to the formation of Cooper pairs with non-zero orbital momenta $l \ne 0$. However, straightforward 
calculations showed that the transition temperature was extremely low 
(the estimate of Kohn and Luttinger \cite{KL}  was $T_c \sim 10^{-40}\,{\rm K}$ for some realistic 
parameters of the fermion system). This  extreme low value of $T_c$ 
was one of the reasons why the effect has not been much
studied in recent years.

In the early nineties, Kagan and collaborators obtained 
a number of interesting results within the Kohn-Luttinger theory \cite{Kagan} (such as cascade transitions, 
Kohn-Luttinger effect in a three dimensional system with long-ranged Coulomb interaction, 
Kohn-Luttinger superconductivity in the Hubbard model, {\em etc.}). 
One of the interesting results was that the 
temperature of the superconducting transition derived in the pioneering paper \cite{KL} was 
shown to be underestimated 
due to the unjustified extrapolation in the expression valid for large orbital momenta
down to the value $l=1$. The transition temperatures calculated in Ref.~\onlinecite{Kagan}
were higher than the original estimate 
but still too low to attract much attention.

One of the most natural issues to be explored has been the status of the Kohn-Luttinger theory
in two dimensions. First of all, the Kohn-Luttinger physics is 
about the formation of bound states.
It is very natural to expect that in the lower dimensionality it is easier to form bound states
({\em i.e.}, Cooper pairs). However, a simple calculation of the polarization operator leads to 
the disappointing result:
no singularity exists in second-order perturbation theory. Namely, the polarization operator 
reads (we use units $\hbar=c=1$ throughout the paper):
\begin{equation}
\label{P0}
\Pi({\bf q}) =
\left\{
\begin{array}{ll}
\nu,  & \mbox{if }\,q < 2 k_{\rm F};\\
 \nu\left[1 -  \sqrt{1 - \left({2 k_F / q} \right)^2}\right], & \mbox{if}\,\,\,q > 2 k_{\rm F},
\end{array}
\right.
\end{equation}
where $\nu = m/ \left(2 \pi \right)$ is the density of states at the Fermi-line, 
$k_{\rm F}$ is the Fermi momentum, and
${\bf q}$ is the momentum transfer in the Cooper channel ($q = 2 p_F \sin{\phi /2}$, where
$\phi$ is the scattering angle).
Let us remember that the attractive harmonics in the scattering amplitude in the three-dimensional
case comes from the well-known logarithmic Kohn's singularity 
$\Pi_{\rm sing}({\phi})= \left( 1 + \cos{\phi} \right)
\ln{ \left( 1 + \cos{\phi} \right)}$ which exists in 3D on both sides of the Fermi-surface. As can be seen
from Eq.(\ref{P0}), the singularity in two dimensions is one-sided which suggests that
no straightforward Kohn-Luttinger effect should exist in 2D. 

In 1993, Chubukov \cite{Chub} showed that this simple scenario was not the complete
story in two dimensions. A two-sided singularity exists, but to find it one
should go beyond second-order perturbation theory. The corresponding transition temperature derived by Chubukov 
reads: $T_c(l) \propto \exp{ \left[-{l^2 / 2  f_0^3}\right]}$, where $f_0$ is the 
dimensionless s-wave scattering amplitude.
Having applied this result to a realistic experiment on $^3{\rm He}-^4{\rm He}$ mixture films, 
the numerical value 
was found as $T_c(l=1)=10^{-4}\,\mbox{K}$.

Let us also mention a recent paper of Guinea {\em et al.} \cite{Guinea} in which
the Kohn-Luttinger physics was phenomenologically incorporated in a model of 
high-$T_c$ cuprates. Within this model the shape of the gap anisotropy has been explored
as a function of doping.

The main idea of the present paper is to search for an effective attractive 
interaction  by taking into account  the frequency dependence of the polarization 
operator, istead of going into higher order perturbation theory. The account for dynamical 
screening, as we shall see below, yields a two-sided singularity. 
Thus, we are looking for a dynamical Kohn-Luttinger effect rather that the original 
static pairing problem as in Refs. [\onlinecite{KL}] and [\onlinecite{Chub}].
Due to the energy dependence of the effective electron-electron coupling, the Cooper problem turns into
an integral equation, similar to the {\'E}liashberg equation in the strong coupling theory 
of superconductivity. \cite{Eliashberg}

Our paper is structured as follows:\\
In Sec. II, we rederive the expression for the polarization operator as a function
of momentum ${\bf q}$ and Matsubara frequency $\omega$. Using this result, we formulate
the Cooper problem and derive the corresponding Bethe-Salpeter equation for the
pairing vertex $\mathcal{T} ({\bf q}; \varepsilon,\varepsilon')$.

In Sec. III, we consider spherical harmonics of the effective interaction 
${\cal V}_l(\omega)$ and show that d-harmonic, which corresponds to the orbital
momentum $l=2$, yields the strongest effective attraction. 

In Sec. IV, we use the explicit expression for the interaction
in the $d$ channel and derive an integral equation for the pairing vertex.
Studying this equation, we show that the pairing vertex may diverge if the 
incoming particles have
high enough energies. We estimate the temperature at which the pairing
with the typical binding energy of $\omega$ commences. We conclude that
at low enough temperature the system is a mixture of low-lying electron
excitations and fluctuating Cooper pairs. We estimate the temperature $T_*$
at which the effect of this fluctuaing pairs becomes essential and may
strongly change transport and thermodynamic properties of the system.

In Sec. V, we briefly discuss the case of long-ranged Coulomb interactions.
We argue that Kohn-Luttinger physics strongly depends on the screening
properties. In a purely two-dimensional system we do not
expect any superconducting instability to survive. If transport is two-dimensional
but screening is three-dimensional, the system is qualitatively described 
by our theory and Ref. [\onlinecite{Chub}].

\section{Cooper problem}

\begin{figure*}
\includegraphics[width=6in]{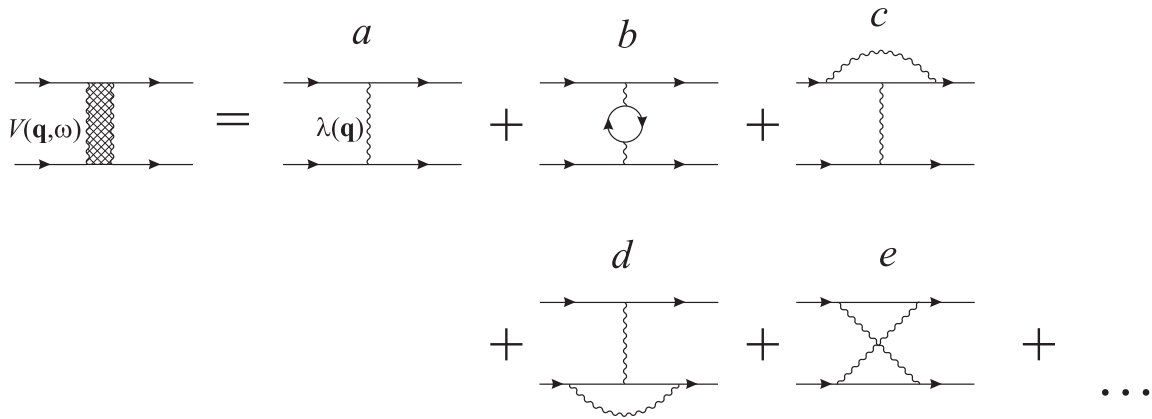}
\caption{\label{fig:fig1} 
Renormalization of the scattering amplitude by Friedel oscillations. Kohn-Luttinger theory.
If the bare coupling is ${\bf q}$-independent diagrams ``b'', ``c'', and ``d'' cancel each 
other out.}
\end{figure*}

Let us start with calculating the effective electron-electron interaction ${\cal V}({\bf q},\omega)$
(where $\omega=2 \pi n T$ is the bosonic Matsubara frequency).
In  second-order perturbation theory there are four diagrams to be considered, which are shown
in Fig. 1. If the bare potential $\lambda({\bf q})$ is short-ranged the diagrams ``b'',``c'', and ``d'' cancel each
other out and only ``e'' contributes to the renormalized interaction. The latter diagram is functionally
identical to ``b'' but depends on $ p + p'$ rather then on $p - p'$, 
where $p = \left({\bf p}, \varepsilon \right)$.
Thus, knowing the two-dimensional polarization operator we readily obtain the total effective electron-electron
coupling. 

The polarization operator is defined as
\begin{eqnarray}
\label{P1}
\mbox{\Large $\pi$} ({\bf q},\omega_m) =
T \sum_{\varepsilon_n}\int
{\displaystyle d ^2 {\bf k} \over \displaystyle  (2 \pi)^2}\,
&&\!\!\!\!\!\!{\cal G}_{\displaystyle\varepsilon_n + {\displaystyle\omega_m \over \displaystyle 2}} \left({\bf k} + 
{ \displaystyle {\bf q} \over \displaystyle 2}\right) \nonumber \\
&& \!\!\!\!\!\!\!\!\!\times {\cal G}_{\displaystyle\varepsilon_n - {\displaystyle \omega_m \over \displaystyle 2}} \left({\bf k} - 
{ \displaystyle {\bf q} \over \displaystyle 2}\right), 
\end{eqnarray}
where ${\cal G}_{\varepsilon}\left({\bf k}\right) = \left( i \varepsilon - \xi_{\bf k} \right)^{-1}$
is the Matsubara Green function, $\xi_{\bf k} = \left( {\bf k}^2 - k_{\rm F}^2 \right) / 2m$, and 
$\varepsilon_n = (2n+1)\pi T$ is the Fermionic Matsubara frequency.
After the sum over $\varepsilon_n$ is evaluated, Eq.(\ref{P1}) takes on the form
\begin{equation}
\label{P2}
\mbox{\Large $\pi$} ({\bf q},\omega_m) =
2\, {\rm Re\,} \left[ \int 
{\displaystyle d ^2 {\bf k} \over \displaystyle  (2 \pi)^2} \, 
 {f({\bf k}) \over \varepsilon({\bf k}) - \varepsilon({\bf k - q}) -
i \omega_m} \right],
\end{equation}
where $f({\bf k})$ is the Fermi distribution. At not very high temperatures $T \ll \varepsilon_{\rm F}$, 
it can be written as
$f({\bf k}) = \theta \left(k_{\rm F} - \left| {\bf k} \right| \right)$ and after a straightforward
calculation we obtain:
\begin{equation}
\label{Pires}
\mbox{\Large $\pi$}(z) = \nu\, {\rm Re\,} 
\left[ 1 - {1 \over {\rm Re\,} z}\, \sqrt{z^2 -1} \right],
\end{equation}
where we have introduced the complex variable $z$ for compactness:
$$
z = {q \over 2 k_{\rm F}} + {i \left| \omega_m \right| \over v_{\rm F} q},
$$
and $v_{\rm F} = k_{\rm F} / m$ is the Fermi-velocity.
One can easily check that Eq.(\ref{P2}) reproduces Eq.(\ref{P0}) if
$\omega_m = 0$. 
To get the expression for the polarization operator $\Pi({\bf q}, \omega)$ as a function of the
real frequency,\cite{Stern} one has to do the analytical continuation in Eq.(\ref{Pires}). 
Let us note here that only the real part of the polarization operator renormalizes the scattering amplitude. 
The imaginary part is not relevant to this renormalization. The latter quantity is proportional to the density 
of the electron-hole pairs.

\begin{figure*}
\includegraphics[width=5in]{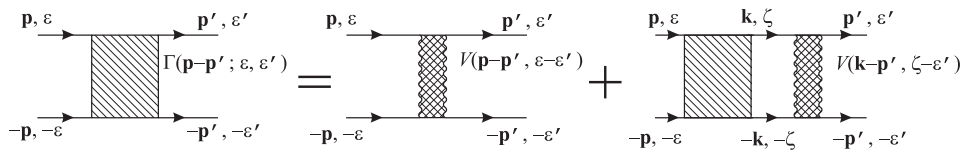}
\caption{\label{fig:fig2}  The Bethe-Salpeter equation for the irreducible vertex 
$\Gamma$ (Cooper problem).}
\end{figure*}

Let us now formulate the Cooper problem for the case under consideration.
We are looking for a singularity in the Cooper channel (see the diagrammatic equation
in Fig.~2). After averaging over the spin indices, the Bethe-Salpeter equation can be
written as:
\begin{eqnarray}
\label{BS}
\mathcal{T} \left( {\bf q}; \varepsilon,\varepsilon'\right) = 
{\cal V} ({\bf q}, \omega)
 - T \sum_{\zeta} \int && \!\!\!\!\!\!\!\!\!{\displaystyle d ^2 {\bf k} \over \displaystyle  (2 \pi)^2} \,
 \mathcal{T} \left( {\bf k - p}; \varepsilon,\zeta \right)   \\
&& \!\!\!\!\!\!\!\!\!\!\!\!\!\!\!\!\!\!\!\!\!\!\!\!\!\!\! \times  {\cal G}_{\zeta}({\bf k}) {\cal G}_{-\zeta} (-{\bf k}) \nonumber
\,{\cal V} ({\bf k - p'}, \zeta - \varepsilon'),
\end{eqnarray}
where $\zeta$, $\varepsilon$, and $\varepsilon'$ are fermionic Matsubara frequencies,
${\bf q} = {\bf p} - {\bf p}'$, and $\omega = \varepsilon - \varepsilon'$.
Let us emphasize that ${\cal V} ({\bf q}, \omega)$ is the renormalized interaction
which depends on momentum and energy transfer.
 
Let us now consider only electrons in the very vicinity of the Fermi surface so that
$q = 2 k_{\rm F} \sin \left(\phi / 2\right)$, where $\phi$ is the scattering angle.
Following the standard route in the Kohn-Luttinger theory, we expand ${\cal V}$ 
and $\mathcal{T}$ in series of the normalized eigenfunctions of the angular momentum:
\begin{equation}
\label{Vex}
{\cal V} ({\bf q}, \omega) = \sum_l {\cal V}_l(\omega) {\Phi_l (\phi)}
\end{equation}
and
\begin{equation}
\label{Gex}
\mathcal{T} \left({\bf q}, ; \varepsilon,\varepsilon' \right) = 
\sum_l \mathcal{T}_l \left( \varepsilon,\varepsilon' \right) {\Phi_l (\phi)},
\end{equation}
where
$$
{\Phi_l (\phi)} = {1 \over \sqrt{2 \pi}}\, {\rm e}^{i l \phi}.
$$
Then, Eq.(\ref{BS}) takes on the form:
\begin{eqnarray}
\label{BS2}
\mathcal{T}_l \left(\varepsilon,\varepsilon'\right) = 
{\cal V}_l (\omega)
- T \sum_{\zeta}
 \mathcal{T}_l \left( \varepsilon,\zeta \right)\, 
 {\cal C}(\zeta)\,
{\cal V}_l (\zeta - \varepsilon'),
\end{eqnarray}
where as usual ${\cal C}(\zeta)$ is the Cooperon, which is the source of
the BCS logarithm:
\begin{equation}
\label{cop}
{\cal C}(\zeta) = \int \left| {\cal G}_{\zeta} \left({\bf k}\right) \right|^2 \, 
{\displaystyle d ^2 {\bf k} \over \displaystyle  (2 \pi)^2} = 
{\pi \nu \over | \zeta |}.
\end{equation}
Let us note that Eq.(\ref{BS2}) is exact at any temperature.
However, we shall consider only the case of low temperatures
to avoid technical difficulties connected with the analytical
continuation in Eq.(\ref{BS2}). 
In the limit $T \to 0$, the procedure of
the analytical continuation reduces to the simple Feynman rotation and all the Matsubara
sums involved may be replaced by the corresponding integrals with the temperature serving 
as a ``low-energy cut-off.'' The main result we are
deriving in the present paper can be noticed in this limit as well.

\section{Effective attraction in the $d$-channel}

The next step is to evaluate the spherical harmonics of the renormalized interaction.
At this point, let us assume that the initial electron-electron interaction is defined only for 
the energies smaller than some threshold value $\tilde{\omega} \ll \varepsilon_{\rm F}$, which will be serving 
as the high-energy cut-off (just as the Debye frequency in the classical weak-coupling BCS theory).
In this case when performing actual calculations we can expand on $\omega/ \varepsilon_{\rm F}$.

The $l$-harmonics of the polarization operator (\ref{Pires}) can be written as:
\begin{equation}
\label{Pil}
\mbox{\Large $\pi$}_l(\omega_m) = \sqrt{2 \over \pi} \int\limits_0^{2 \pi} 
\mbox{\Large $\pi$}(\phi,\omega_m) \cos{l \phi}\, d\phi.
\end{equation}
Keeping in mind that $\omega \ll \varepsilon_{\rm F}$ and evaluating the integral
with the logarithmic accuracy, we obtain for even orbital momenta $l=2n$:
\begin{eqnarray}
\label{Peven}
\mbox{\Large $\pi$}_{2 n}(\omega) = - \sqrt{2 \over \pi} \nu\, 
{\left| \omega \right| \over 2 \varepsilon_{\rm F}} \Biggl\{
&& \!\!\!\!\!\!\! {3 \over 2} \ln{2 \varepsilon_{\rm F} \over \left| \omega \right|}  \\
&& \!\!\!\!\!\! - 2
\left[ \psi\left(n + {1 \over 2}\right) - \psi\left({1 \over 2}\right) \right] 
\Biggr\}\nonumber
\end{eqnarray}
and for $l=2n+1$
\begin{eqnarray}
\label{Podd}
\mbox{\Large $\pi$}_{2 n+1}(\omega) = -\sqrt{2 \over \pi} \nu\, 
{\left| \omega \right| \over 2 \varepsilon_{\rm F}}
\Biggl\{
&& \!\!\!\!\!\!\!
{1 \over 2} \ln{2 \varepsilon_{\rm F} \over \left| \omega \right|} \\
&& \!\!\!\!\!\! - 
2 n\, {\left| \omega \right| \over 2 \varepsilon_{\rm F}}
\left[ {\bf C} + \psi\left(n + 1 \right)\right] \Biggr\} \nonumber,
\end{eqnarray}
where $\psi$ is the logarithmic derivative of the Gamma-function and ${\bf C} \approx 0.577$ is
the Euler's constant. 

From Eqs.(\ref{Peven},\ref{Podd}) we see that the dependence on the orbital momentum is very weak.
The effective interaction can be written as
\begin{equation}
\label{Vl}
{\cal V}_l(\omega) =  \mbox{\Large $\pi$}_l(\omega)
\left\{ \lambda(0)\left( -1 \right) ^{-l} +  2 \left[
\lambda(0)\lambda(2 k_{\rm F}) - \lambda^2(2 k_{\rm F})
\right] \right\}, 
\end{equation}
where $\lambda({\bf q})$ is the Fourier-component of the bare interaction potential.

If the initial interaction is ${\bf q}$-independent, we see that
the effective interaction is attractive only for the even values
of the orbital momentum $l=2n \ne 0$. The effective attraction
is the strongest for $l=2$. The corresponding $d$-harmonics 
reads: \cite{note}
\begin{equation}
\label{Vd}
{\cal V}_{d}(\omega) = - {3 \over \sqrt{2 \pi}}\, \nu \lambda^2\,
{\left| \omega \right| \over 2 \varepsilon_{\rm F}}
\ln{ {2 \varepsilon_{\rm F} \over \left| \omega \right|}}.
\end{equation}

\section{Pairing at finite energies}

We can substitute  result (\ref{Vd}) into the Bethe-Salpeter equation 
(\ref{BS2}) which turns into an integral equation (at $T \to 0$) with
a well defined kernel 
$K( \varepsilon,  \varepsilon') = {\cal V}_{d}(\varepsilon -  \varepsilon')
{\cal C}(\varepsilon')$. One can easily see that if the incoming particles
have zero energies, the Cooper singularity gets canceled. However, 
at finite energies the Cooper logarithm survives being cut-off
by the energy transfer.

For further treatment, let us define the following auxiliary dimensionless variables and functions:
$$
x = \varepsilon / 2 \varepsilon_{\rm F},
$$
$$
\tilde{x}= \tilde\omega / \varepsilon_{\rm F},
$$
$$
g_0(x) = - \left| x \right| \ln{1 \over \left| x \right|},
$$
$$
g(x,x') = \left[ {3 \over \sqrt{2 \pi}}\, \nu \lambda^2 \right]^{-1} {\cal T} 
(\varepsilon,  \varepsilon'),
$$
and
$$
\kappa = {3 \pi \over \left( 2 \pi \right) ^{3/2}} \left( \lambda \nu \right)^2.
$$
In these notations, Eq.(\ref{BS2}) takes on the form
\begin{equation}
\label{BSformal}
g(x,x') = -g_0(x - x') + \kappa \int dy\, {g_0(x - y) \over \left| y \right|}\,
g(y,x').
\end{equation}
The integral in (\ref{BSformal}) is defined in such a way that the large-$y$ singularities
are cut-off by $\tilde\omega / \varepsilon_{\rm F}$ and low-$y$ singularities at
$\tau = T / 2 \varepsilon_{\rm F}$. 

It is hard to solve Eq.(\ref{BSformal}) exactly.
However, we are mostly interested not in the detailed solution but in the possibility
of a singularity in the pairing vertex $g(x,x')$ which would be a signal
of a superconducting pairing (but not necessarily a global superconducting instability). 
Let us emphasize here that $g(0,0)=0$ by the construction
and it can not diverge simply because there is no attraction in this case, unless we take
into account the higher order diagrams. At finite energy transfers, the large Cooper logarithm
appears which yields a divergence of $g(x,x')$ which we interpret as an appearance
of fluctuating Cooper pairs built up of the electronic excitations with finite energies.
One of the ways to search for the singularity is to consider the eigenvalue problem
for the  kernel of integral equation (\ref{BSformal}):
\begin{equation}
\label{Delta}
\Delta(x) = \kappa \int {\left| x - y \right| \over \left| y \right|}\, 
\ln {1 \over \left| x - y \right|}
\Delta(y) dy.
\end{equation}
The singularity exists if there is a non-trivial solution of this equation.
To get some qualitative estimates let us approximate the corresponding eigenvector
by the following trial function:
\begin{equation}
\label{trial}
\Delta(x) = \Delta_0 + \Delta_1 \left| x \right|,
\end{equation}
where $\Delta_0$ and $\Delta_1$ are some weak (logarithmic) functions of $x$.
From Eqs.(\ref{Delta},\ref{trial}) we can derive the self-consistency equation
which yields the estimate for the threshold temperature at which the pairing with the typical 
energy transfer of $\omega$ commences:\cite{note2}  
\begin{equation}
\label{Tp}
T_p(\omega) \sim \omega\, \exp{
\left\{-{ 1 \over k^2 \tilde{x}^2 \ln{\left( 2 \varepsilon_{\rm F} / \omega \right)}}\right\}}.
\end{equation} 
This estimate can be alternatively derived by considering the resolvent
of the integral equation straightforwardly. Namely, one can formally re-write Eq.(\ref{BSformal})
as follows:
$$
\hat{g} = g_0 + \kappa\, \hat{K}\, \hat{g},
$$
where $\hat{K}$ is the operator with the kernel in $x$-repre\-sen\-ta\-tion being
equal to $K(x,y) = {\left| x - y \right| \over \left| y \right|}\, 
\ln {1 \over \left| x - y \right|}$.
The solution of this equation has can be formally written as:
$$
\hat{g} = \hat{R}(\kappa)\,  g_0 = \left[1 - \kappa \hat{K} \right]^{-1}\, g_0,
$$
where $\hat{R}(\kappa)$ is the resolvent, which can be also written as:
\begin{equation}
\label{res}
\hat{R}(\kappa) = \sum\limits_{n=0}^{\infty} \kappa^n \hat{K}^n,
\end{equation}
where $\hat{K}^n$ can be found by evaluating the convolution
of the corresponding kernels in the $x$-representation:
$$
K^{(n)}(x,y) = \int K(x,z) K^{(n-1)}(z,y) dz.
$$
Studying the geometric series (\ref{res}), one can see that its $2n$'s term contains the logarithm
$\ln^{n}{\left( \omega / T \right)}$, with $\omega$ being the typical energy of the electrons
in the Cooper channel. Summing up the series, we reproduce Eq.(\ref{Tp}).

The integral equation (\ref{BSformal}) and the corresponding eigenproblem (\ref{Delta})
are mathematically well-defined for any $x \lesssim \tau$ ({\em i.e.} $\omega \lesssim T$). However, it does not
make too much sense to study the structure of the solutions at such energies in the framework
of our formalism based on the Matsubara technique. Thus, result (\ref{Tp}) has the following domain
of applicability:
$$T \ll \omega \lesssim \tilde\omega \ll \varepsilon_{\rm F}.$$
Working in this domain, the replacement of the Matsubara sums by the integrals is
legitimate and our interpretation of $\omega \gg T$ as a real energy of a pair
is valid as well. 

\begin{figure}
\includegraphics[width=1.5in]{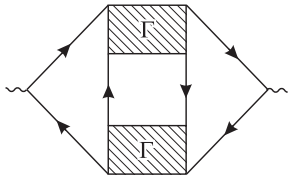}
\caption{\label{fig:fig3} Aslamazov-Larkin contribution to the conductivity. 
Small wavy lines correspond to the 
factor $e {\rm v}$. Shaded boxes are the pairing vertexes which in the case under
consideration are functions of the four variables $\Gamma(k_1,k_2;k_1',k_2')$.}
\end{figure}
Let us now briefly discuss how the appearance of the fluctuating pairs affects
the physical properties of the system.
The correction to the conductivity  is described by the diagrams similar
to the ones in the conventional fluctuation theory \cite{AL} (see {\em e.g.} Fig.~3 where the 
Asla\-ma\-zov-Lar\-kin-like diagram is shown). It is a rather difficult problem
to calcu\-late the corre\-s\-ponding contri\-bu\-tions in the case under consideration. 
However, we can get some qualitative insight by noting that the analytical continuation 
to the real frequencies in the expression for the conductivity contains the factor
$\coth{\left( \omega \over 2 T \right)}$, which is basically the Bose-distribution
for the fluctuating Cooper pairs (the density of the Cooper pairs). This factor
and the corresponding correction are exponentially small unless there exist
Cooper pairs with $\omega \sim T$. Using Eq.(\ref{Tp}), we can estimate
the temperature $T_*$ at which such pairs appear. It is defined by the condition
$T_p(T_*) \sim T_*$. Thus, we readily obtain:\cite{note2}
\begin{equation}
\label{T*}
T_* \sim \varepsilon_F \, \exp{ \left\{ - 
\left[ {\left(2 \pi \right)^{3/2} \over 3 \pi}\, {\varepsilon_{\rm F} \over \tilde\omega} \right]\,
{1 \over \left( \lambda \nu \right)^4} \right\}}.
\end{equation}
At this temperature, contribution to the conductivity due to the preformed
Cooper pairs may become comparable to the Drude conductivity of a normal metal. 

\section{Long-range Coulomb interaction}

Until now, we have been studying a Fermi-system with the bare electron-electron
coupling being short-ranged. It is worth considering the case when the initial 
interaction is the long-range Coulomb repulsion. In this case our treatment
is not applicable since the momentum-dependence of the Coulomb interaction
becomes crucial. However, we can get some qualitative insight into the problem
without cumbersome calculations. There are several possibilities one can consider:

First, we can study a system in which both transport
and screening are two-dimensional. In this case, we can readily conclude that
there is no possibility for Kohn-Luttinger pairing because the 
long-wavelength Thomas-Fermi screening
is weak 
\begin{equation}
\label{C1}
V({\bf r}) =  \int {d^2 {\bf q} \over \left(2 \pi \right)^2} \,
{2 \pi e^2 \over q} {1 \over \epsilon(q)} {\rm e}^{i {\bf q} {\bf r}} \propto
{1 \over r^3}, 
\end{equation}
where $a_0 = 1 / m e^2$ is the effective 2D screening length 
and the Thomas-Fermi dielectric function has the standard long-wavelength form:
\begin{equation}
\label{C2}
\epsilon(q) = 1 + {2 \over a_0 q}.
\end{equation}
We can now calculate the spherical harmonics of the screened 
Coulomb interaction
\begin{equation}
\label{C3}
V_l = \sqrt{2 \over \pi} \int {2 \pi e^2 \over q + 2/a_0}\, \cos{l\phi}\, d\phi, \,\,\, q=2k_{\rm F} \sin{\phi \over 2},
\end{equation} 
which are certainly all repulsive and remain repulsive even after Friedel oscillations
are taken into account (see Fig.1). Thus, even going beyond this long-wavelength Thomas-Fermi
analysis we do not expect any pairing instability for the long-ranged
Coulomb interaction to appear, as long as both transport and screening are two-dimensional.
However, the account for the dynamically screened Coulomb interaction may lead to other important
effects such as renormalization of the Fermi-liquid parameters (effective mass, $g$-factor, {\em etc.}).
This issue is currently being investigated by the authors and the results will be reported elsewhere.

Second, one can consider a system in which transport is two-dimensional but screening
is three-dimensional. In this case the Coulomb interaction is well screened and 
decays exponentially at large distances $V(r) \propto \exp{\left(-r/d\right)}$ ($d$ is
the screening length). In the limit $k_{\rm F} d \ll 1$,  the potential becomes
effectively short ranged and, thus, the theory developed in the present paper is qualitatively 
valid. Let us note that in this model
the high-energy cut-off is basically the Fermi energy which  violates the assumption
$\tilde\omega \ll \varepsilon$ we used in our calculations. This, however, should not
change main qualitative result of the paper.

There is also an intermediate situation which may exist when the two-dimensional Fermi-liquid
lives in the very close vicinity of a metallic substrate. In this situation, each two-dimensional
electron produces an image in the metallic substrate so that the bare electron-electron
interaction decays only as $r^{-3}$ at large distances. In this case, there is no simple answer
whether the Kohn-Luttinger pairing exists or not. Presumably, the Kohn-Luttinger pairing in such
a setup is possible if the Fermi-liquid is dilute enough, so that Friedel oscillations
may compete with the initial dipole-dipole coupling.

\section{Conclusion}

Before concluding, we point out that earlier theoretical work in the literature has considered 
\cite{R} the possibility of bound states and Cooper pairing in a dilute 2D system of 
fermions interacting via a short-ranged repulsive interaction.  Engelbrecht and Randeria
have considered a regular expansion in the $T$-matrix in two-dimensions analogous to
the expansion on the dilute gas parameter $k_{\rm F} a \ll 1$ in 3D.\cite{Gal}
Apart from the three-dimensional result, they have found an unusual pole in the particle-particle
channel.
Although we do not find
any obvious connection between our microscopic analysis and this earlier work, \cite{R}
the claim of a new 2D collective mode interpreted as a bound excitation of two holes
is somewhat reminiscent of our finding in this paper that Kohn-Luttinger type superconducting
pairing is possible at finite excitation energies. Whether there is a deep connection
between our work and the earlier results \cite{R} remains unclear at this stage.

Summarizing, we have shown that a clean two-\-dimen\-si\-onal Fermi-system with a short-range
repulsive interaction between electrons becomes unstable against a formation
of $d$-wave Cooper pairs with a finite binding energy at a low enough temperature. 
Thus, the low-temperature state of the system is a mixture of low-lying electron excitations and 
preformed fluctuating Cooper pairs. The new  type of carriers may noticeably change 
the physical properties of the system such as conductivity, susceptibility, {\em etc.}
at a temperature $T_*$ [see Eq.(\ref{T*})]. Let us note that from our theory
it follows that the fluctuating pairs appear within the normal state having
a finite gap which is connected with the binding energy.
Note that there is no global superconductivity specifically predicted in our
theory, only a pseudo-pairing at finite excitation energies.
The results obtained in the present paper may be relevant to the pseudogap
experiments in high-$T_c$ superconductors. \cite{pseudo}

This work was supported by the US-ONR, the LPS, and DARPA. V.G. wishes to thank
A. I. Larkin, M. Yu. Kagan, and M. Pustil'nik for useful discussions and A. Yu. Kaminski for the help
in preparation of the manuscript.


\begin{thebibliography}{}

\bibitem{KL} W. Kohn and J. H. Luttinger, Phys. Rev. Lett. {\bf 15}, 524 (1965).

\bibitem{Kagan} M. A. Baranov, A. V. Chubukov, and M. Yu. Kagan, Int. J.
Mod. Phys. B {\bf 6}, 2471 (1992); M. Yu. Kagan, P. Brussaard, and H. W. Capel,
Phys. Lett. A {\bf 221}, 407 (1996); M. A. Baranov and M. Yu. Kagan,
Z. Phys. B {\bf 86}, 237 (1992).

\bibitem{Chub} A. V. Chubukov, Phys. Rev. B {\bf 48}, 1097 (1993). 

\bibitem{Guinea} F. Guinea, R. S. Markiewicz, and M. A. H. Vozmediano, cond-mat/0206208.

\bibitem{Eliashberg} G. M. \'{E}liashberg, Sov. Phys. JETP {\bf 11}, 696 (1960).

\bibitem{Stern} F. Stern, Phys. Rev. Lett. {\bf 18}, 546 (1967)

\bibitem{note2} The approximation we use allows us to determine the exponents in 
Eqs.(\protect\ref{Tp},\protect\ref{T*}) up to a numerical factor only. 

\bibitem{note} Stricktly speaking in Eq.(\protect\ref{Vd}), we have to replace the bare interaction $\lambda$
with the full $s$-wave scattering amplitude $f_0$. In the leading approximation
the renormalized amplitude is $f_0 = \left[ \left( \lambda \nu \right) ^{-1} + 
\ln{\left( a k_{\rm F} \right)^{-2}} \right]^{-1}$, where $a$ is the range of the 
potential.

\bibitem{AL} L. G. Aslamazov and A. I. Larkin, Sov. Phys. Solid State {\bf 10}, 875
(1968); K. Maki Progr. Teor. Phys. {\bf 40}, 193 (1968); R. S. Thomson Phys. Rev. B
{\bf 1}, 327 (1970).

\bibitem{R} J. R. Engelbrecht and M. Randeria, Phys. Rev. Lett. {\bf 65},
1032 (1996); J. R. Engelbrecht and M. Randeria, Phys. Rev. B {\bf 45}, 12419 (1992).

\bibitem{Gal} V. M. Galitskii, Zh. Eksp. Teor. Fiz. {\bf 34}, 151
(1958) [Sov. Phys. JETP {\bf 7}, 104 (1958)]; see also: 
K. Huang and C. N. Yang, Phys. Rev  {\bf 105}, 767 (1957);
T. D. Lee and C. N. Yang, {\em ibid.} {\bf 105}, 1119 (1957). 

\bibitem{pseudo} T. Timusk and B. Statt, Rep. Prog. Phys. {\bf 62}, 61 (1999).


\end{thebibliography}
\end{document}